\documentclass[a4paper,11pt]{article}

\usepackage{jinstpub} 
\usepackage{color,soul}

\usepackage{verbatim}
\usepackage{lineno}
\usepackage[utf8]{inputenc}
\usepackage{graphicx}
\usepackage{subcaption}
\graphicspath{{./images/}}

\newsavebox{\subcapboxA}
\newsavebox{\subcapboxB}
\newsavebox{\subcapboxC}

\title{\boldmath A Low-Power 1~Gb/s Line Driver with Configurable Pre-Emphasis for Lossy Transmission Lines}

\author[1]{N. St. John,\note{Corresponding author.}}
\author[1]{S. Mandal,}
\author[1]{G. W. Deptuch,}
\author[1]{E. Raguzin,}
\author[1]{S. Rescia}

\affiliation[1]{Brookhaven National Laboratory, Upton, NY, USA}

\emailAdd{nstjohn@bnl.gov}

\abstract{A line driver with configurable pre-emphasis is implemented in a 65~nm CMOS process. The driver utilizes a three-tap feed-forward equalization (FFE) architecture. The relative delays between the taps are selectable in increments of 1/16th of the unit interval (UI) via an 8-stage delay-locked loop (DLL) and digital interpolator. It is also possible to control the output amplitude and source impedance for each tap via a programmable array of eight source-series terminated (SST) drivers. The entire design consumes 9~mW from a 1.2~V supply at 1~Gb/s.}

\keywords{Digital electronic circuits, Digital signal processing (DSP), VLSI circuits}

\proceeding{Topical Workshop on Electronics for Particle Physics (TWEPP)\\
  19-23 September, 2022\\
  Bergen, Norway}

\begin{document}
\maketitle
\flushbottom

\section{Introduction}
\label{sec:intro}
Modern front-end ASICs require line drivers to send the digitized measurement data off-chip. Such line drivers must be optimized for the specific transmission line (i.e., cable) used by the system to ensure a low bit error rate (BER) while also minimizing power consumption. Furthermore, as data rates continue to increase, transmission line effects (signal attenuation and reflections) begin to hamper the performance of the data link. In particular, attenuation generally increases with frequency, resulting a transfer function (TF) for the transmission line that is low-pass in nature, while reflections give rise to sharp maxima/minima at particular frequencies due to wave interference. In order for transmitters to counteract the frequency-dependent attenuation of the cable, which generates pulse distortion and inter-symbol interference (ISI) in the time domain, designers often utilize a technique known as pre-emphasis or feed-forward equaliaztion (FFE)~\cite{bulzacchelli2015equalization}.

Pre-emphasis is a technique in which the BER for a given data rate can be improved via controlled pre-distortion of the output data signal, in particular by adding overshoot (``emphasis'') around data transitions~\cite{casall_2001}. This process adds high-frequency components to the signal. Ideally, one would want to have an output waveform with a TF that is the inverse of the cable in order to completely cancel the low-pass TF of the cable over a desired frequency range.

\begin{figure}
    \begin{subfigure}{0.33\textwidth}
        \centering
        \includegraphics[width=5cm]{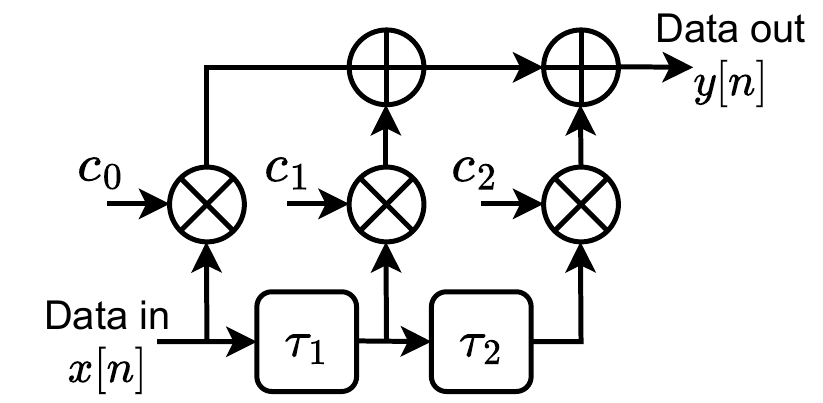}
        \caption{}
    \end{subfigure}
        \begin{subfigure}{0.33\textwidth}
        \centering
        \includegraphics[width=5cm]{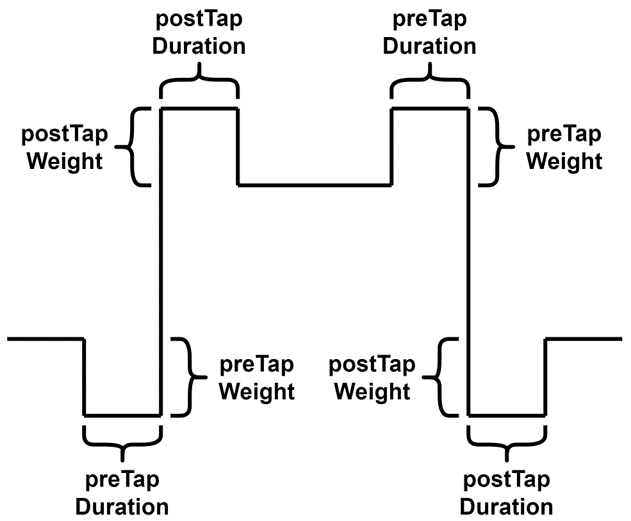}
        \caption{}
    \end{subfigure}
    \begin{subfigure}{0.33\textwidth}
        \centering
        \includegraphics[width=5cm]{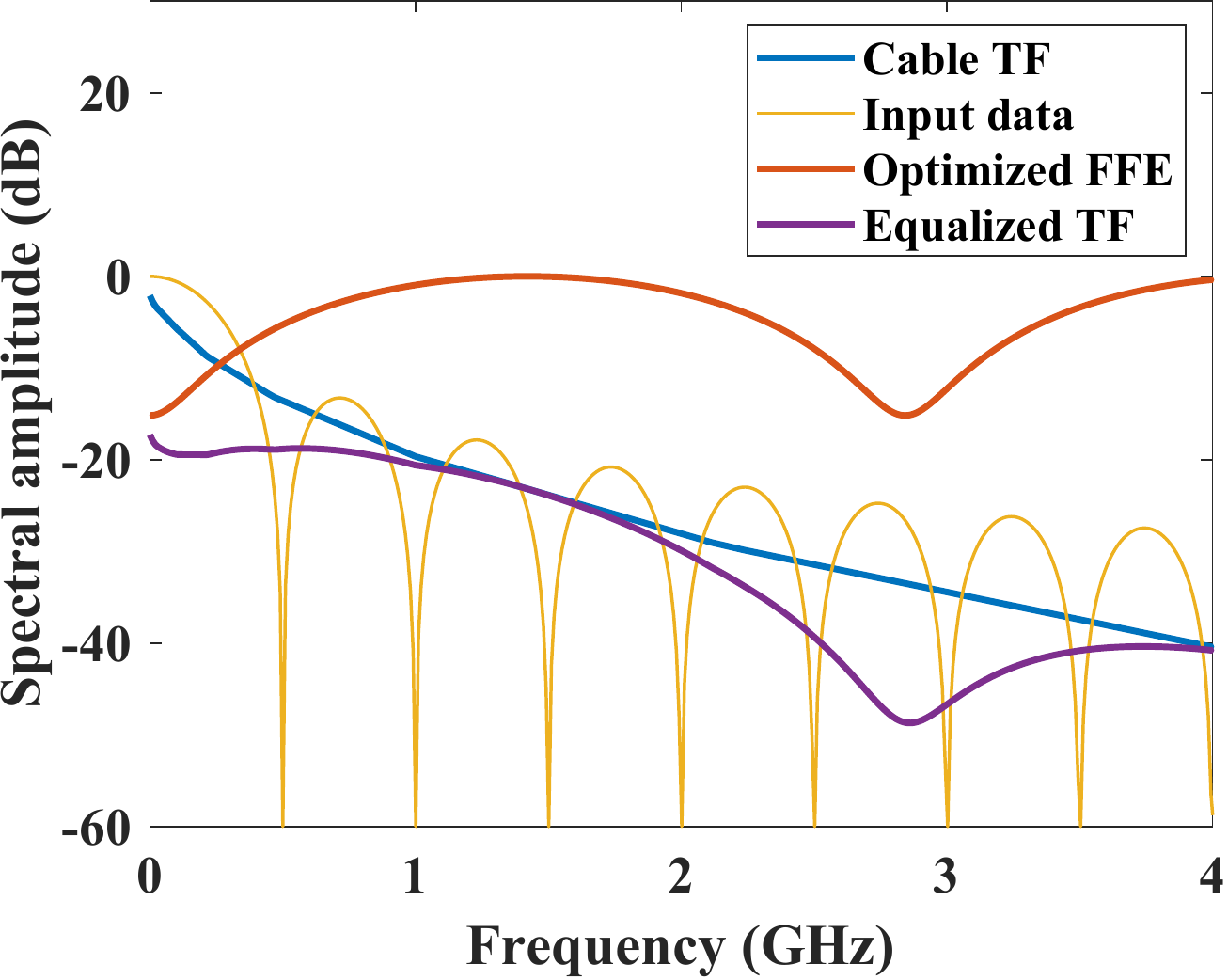}
        \caption{}
    \end{subfigure}
    \caption{(a) Block diagram of the FFE architecture. (b) General pre-emphasized output waveform. (c) Simulated TF of a 3~m length of radio-pure planar transmission line on a flexible Taiflex$^{\mathrm{TM}}$ substrate compared with the power spectrum of random digital data at 0.5~Gb/s. The optimized TF of a three-tap FFE and the resulting equalized TF of the channel are also shown.}
    \label{fig:theory}
\end{figure}

Pre-emphasis is implemented via an $N$-tap finite impulse response (FIR) filter as shown in Fig.~\ref{fig:theory}(a). The example shown in the figure uses three taps ($N=3$) with programmable inter-tap delays, $\tau_{i}$, and tap weights, $c_{i}$. The resulting continuous-time transfer function is given by $H(f) = c_{0}+\sum_{i=1}^{N}{c_{i}\exp\left(-i2\pi f\sum_{j=1}^{i}{\tau_{j}}\right)}$ and should have a high-pass response. Conventional FFE implementations constrain the delays to $\tau_{i}=1/f_{clk}=1\times UI$ where $f_{clk}$ is the clock frequency and $1/f_{clk}$ is the unit interval (UI)~\cite{10.1145/3018009.3018048}. In this case, $H(f)$ is periodic with a period of $f_{clk}$, so the desired high-pass response is only obtained over a limited frequency range (up to $1/(2f_{clk})$). In addition, the output voltage swing is limited by power supply headroom, so the FFE coefficients must be normalized such that $\sum_{i=0}^{N}|c_{i}|=1$. Thus, adding extra taps may not significantly improve performance because the weights of the other taps must decrease to satisfy the headroom constraint, which limits the peak gain. Additionally, the shunt capacitance of the driver increases approximately linearly with the number of taps, thus reducing its output bandwidth. Thus, we restrict ourselves to a three-tap design ($N=3$). In addition, we relax the constraint that all the $\tau_{i}$'s are equal to $1\times UI$, thus allowing for greater flexibility in synthesizing the desired TF over a broad frequency range. This property is particularly useful in compensating for the attenuation of high-loss cables, such as radio-pure microstrip lines on flexible Taiflex$^{\mathrm{TM}}$ substrates developed for the nEXO experiment.

The generic shape of the FFE output waveform is shown in Fig.~\ref{fig:theory}(b). The weight and duration of the first tap (known as preTap) and the third tap (known as postTap) define the amount of pre-emphasis and are optimized to minimize BER (i.e., maximize the eye opening) for a given cable and data rate. As an example, we used electromagnetic simulations to extract a broadband model of a 3~m long radio-pure cable. The results, which are shown in Fig.~\ref{fig:theory}(c), reveal a loss of $-7$~dB/m at 1~GHz. The FFE was optimized to cancel this frequency-dependent channel loss over a bandwidth of 1.5~GHz, which is adequate for our assumed data rate of 0.5~Gb/s (i.e., $UI=2$~ns). The equalized TF, which is also shown in the figure, is nearly flat over this bandwidth, as desired. By contrast, note that a conventional FFE (with delays limited to $1\times UI$) would only be able to equalize the TF up to a frequency of $0.5/UI = 0.25$~GHz.

The rest of this paper is organized as follows. The architecture of the line driver is described in Section~2. Section~3 discusses the scheme used to optimize the output waveform, while Section~4 describes a test chip designed in a 65~nm process. Section~5 discusses both simulated and experimental results of the optimized line driver design, while Section~6 concludes the paper.

\section{Circuit Design}
\begin{figure}
    \centering
    \includegraphics[width=0.74\textwidth]{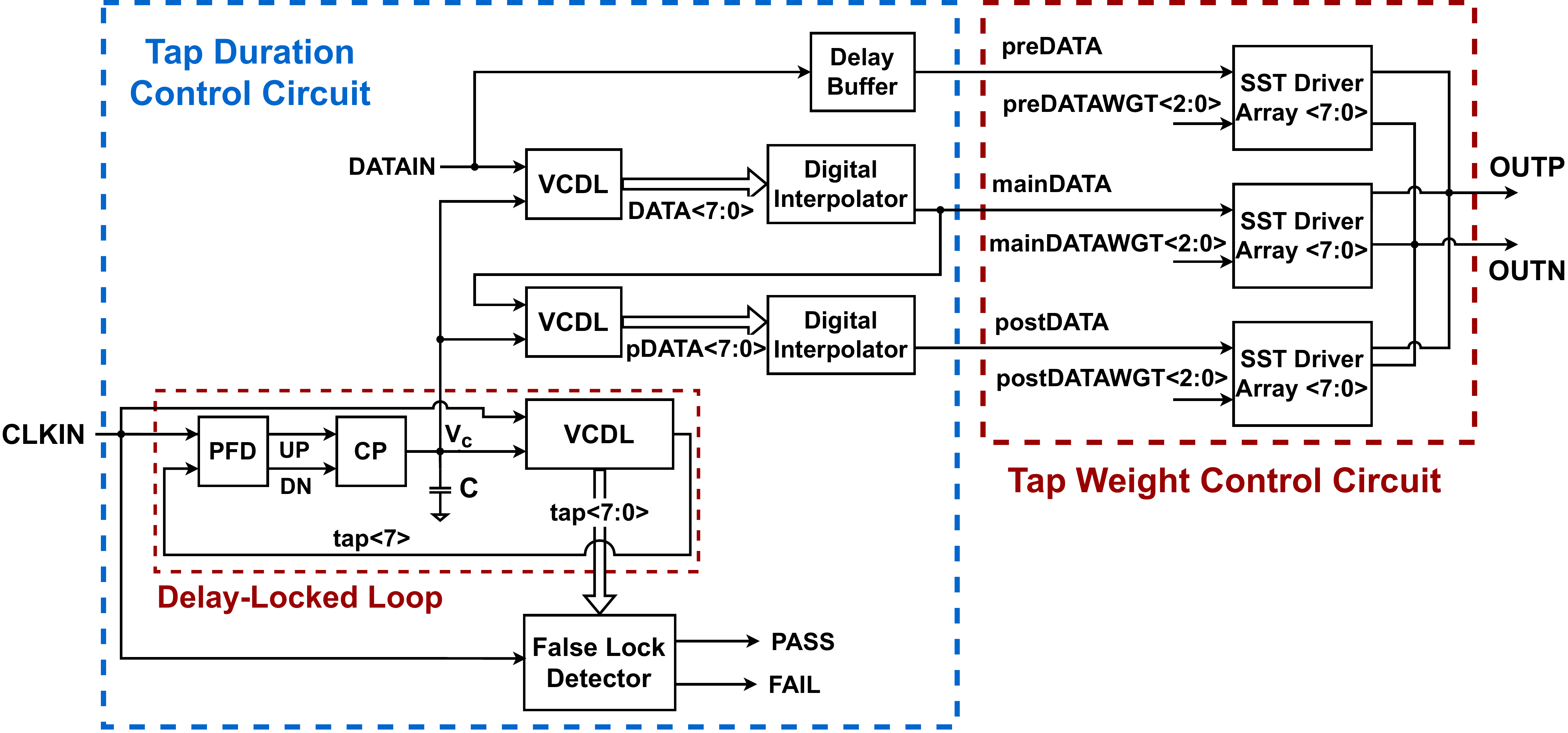}
    \includegraphics[width=0.24\textwidth]{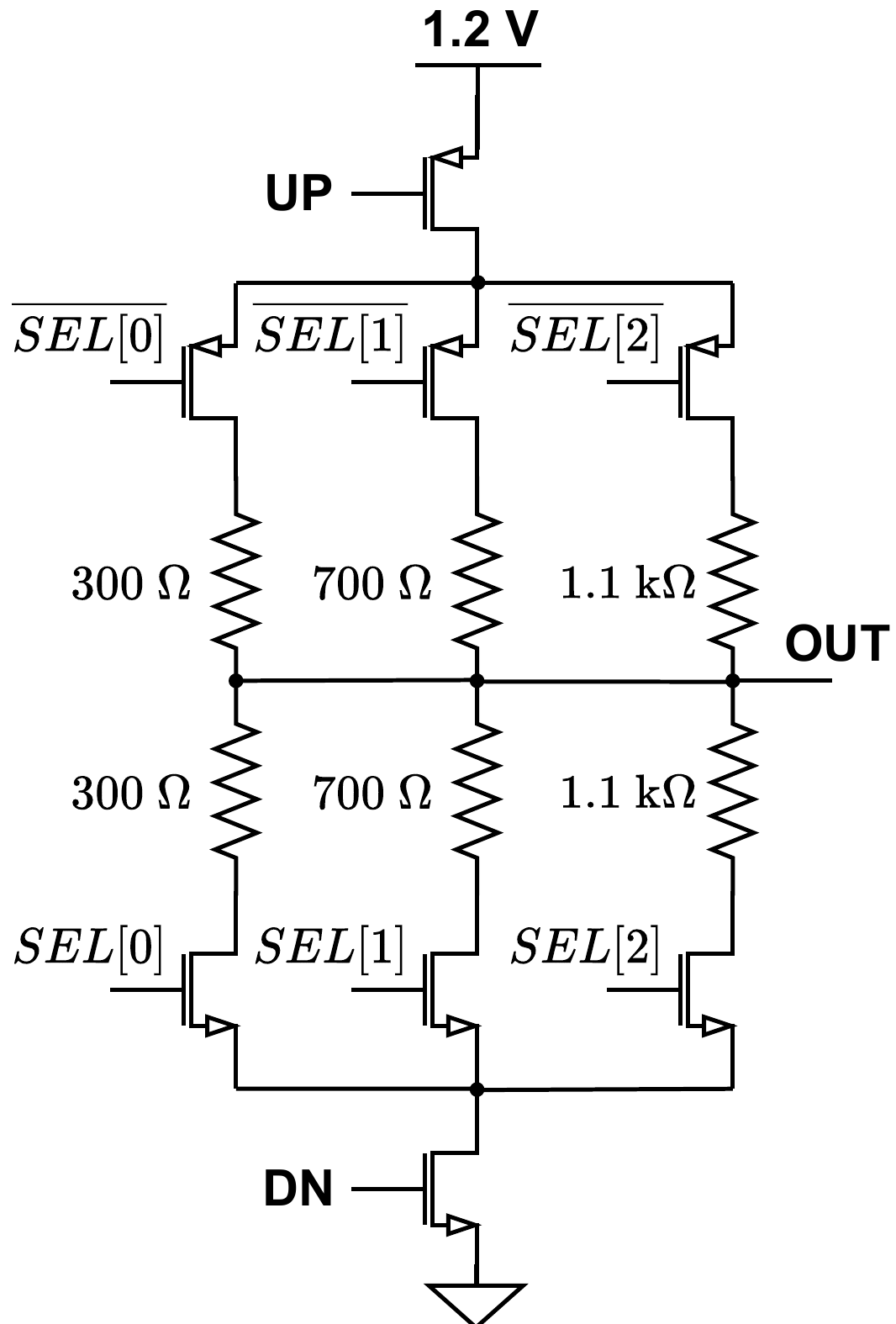}
    \caption{(a) Block diagram of the proposed line driver. (b) Schematic of an SST driver with configurable source resistance; three arrays of such drivers are used to generate the output waveform.}
    \label{fig:overview}
\end{figure}

A block diagram of the line driver is shown in Fig.~\ref{fig:overview}. The design consists of two main parts, namely the tap duration control and tap weight control circuits. These work in tandem to generate the proposed fully-configurable output waveform and are discussed below.  

\textbf{Tap Duration Control Circuitry:}~We utilize a multi-tap delay-locked loop (DLL) to generate user-configurable tap durations that are robust to process, voltage, and temperature (PVT) variations. Specifically, we use an 8-stage DLL with a standard architecture consisting of a phase-frequency detector (PFD), charge pump (CP), and voltage-controlled delay line (VCDL). A capacitive load at the output of the CP creates a control voltage, $V_c$, that stabilizes upon locking of the DLL. A novel single-test false lock detector also monitors the DLL to ensure proper operation, as discussed later. The control voltage $V_c$ then feeds two standalone 8-stage replica VCDLs that generate the delays of the main and post-cursor taps. Each creates eight delayed copies of its input waveform with relative delays equal to those of the stages in the main VCDL. The delayed data waveforms then feed a single-stage digital interpolator (DI)~\cite{garlepp1999portable} that allows the user to interpolate between adjacent taps, thus increasing the timing resolution by $2\times$. Thus, there are 16 uniformly-spaced relative delay options for each tap, resulting in a resolution of $UI/16$. Finally, a buffer that accounts for the delay of the DI inputs the system data and outputs the pre-cursor tap. This concept of delaying the data rather than re-timing it via flip-flops is ideal for high-speed design as it does not require power-hungry high-speed flip-flops. Furthermore, the use of replica VCDLs allows designers to easily add additional taps by daisy-chaining them, as we have done here.

\begin{figure}
\sbox{\subcapboxA}{%
  \begin{subfigure}{0.58\textwidth}
  \includegraphics[width=\textwidth]{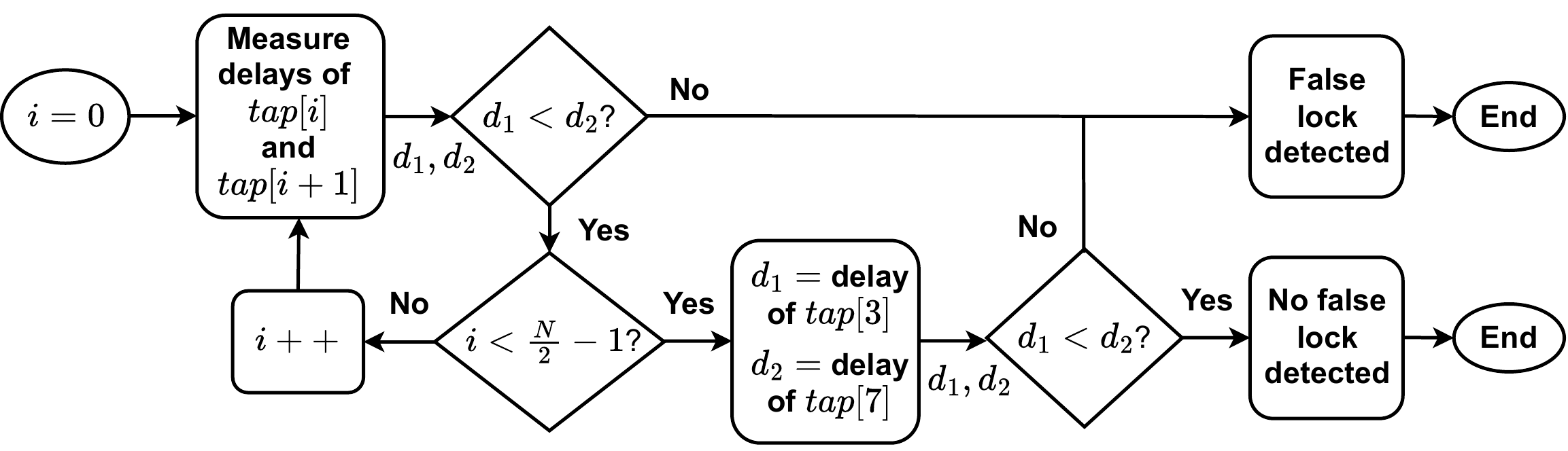}
  \caption{}
  \end{subfigure}%
}
\sbox{\subcapboxB}{%
  \begin{subfigure}{0.38\textwidth}
  \includegraphics[width=\textwidth,height=1.3\textwidth]{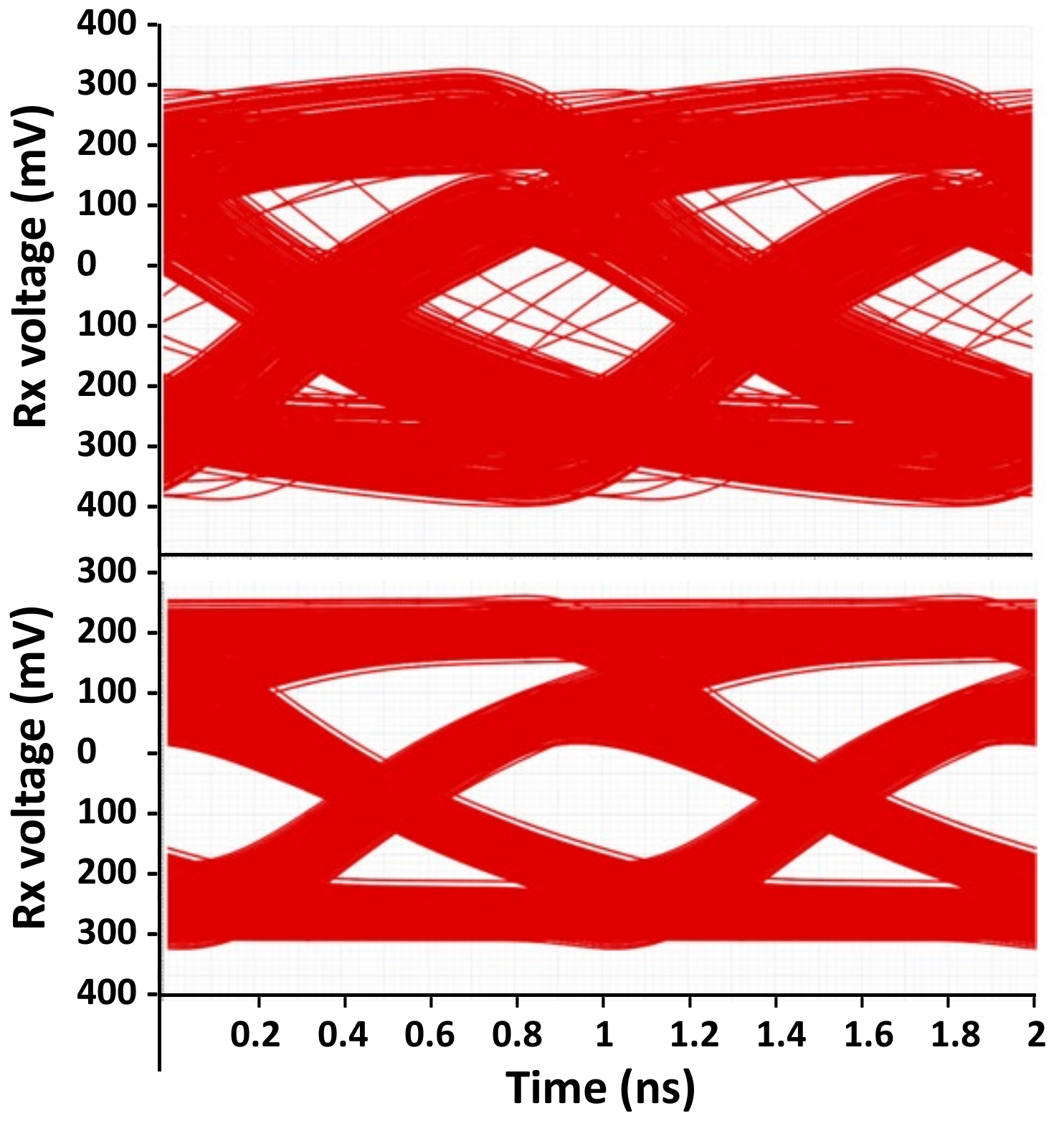}
  \caption{}
  \end{subfigure}%
}
\sbox{\subcapboxC}{%
  \begin{subfigure}{0.58\textwidth}
  \includegraphics[width=\textwidth]{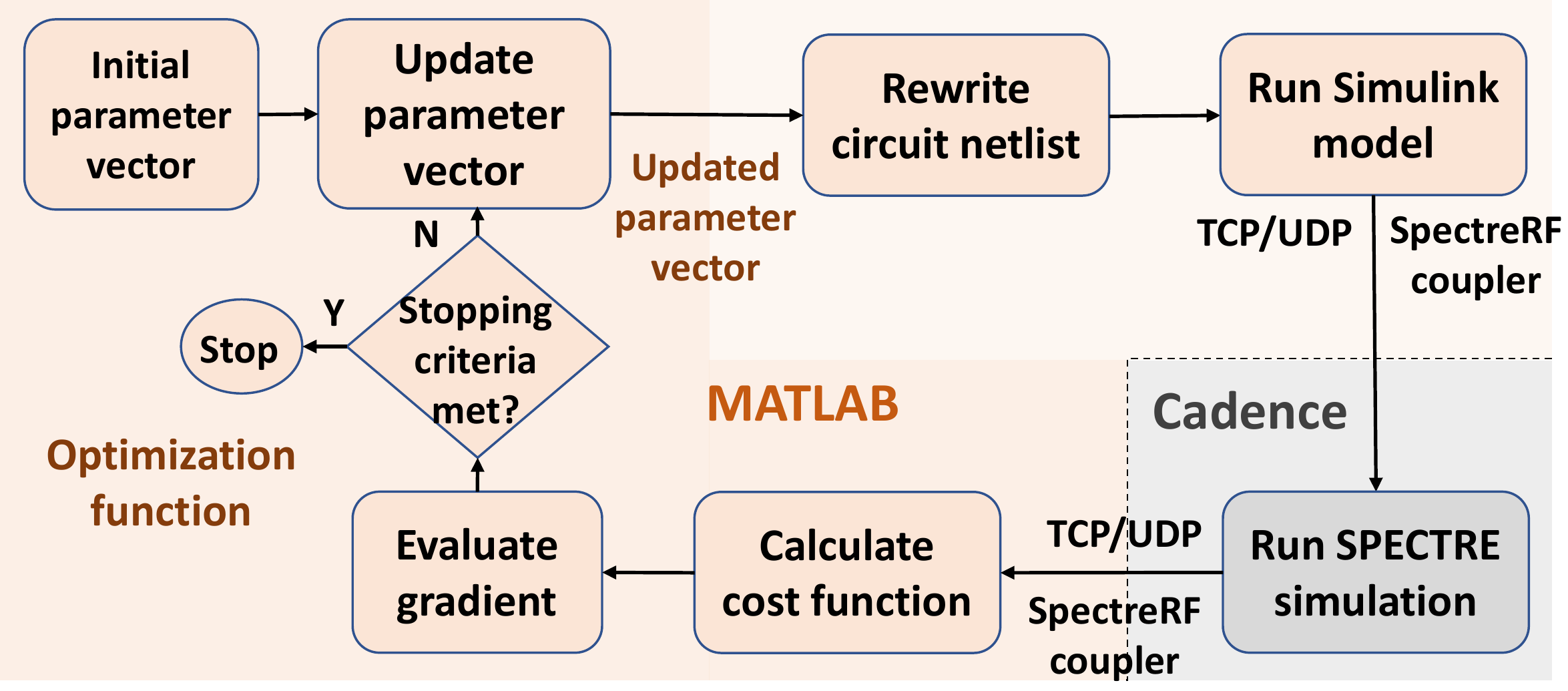}
  \caption{}
  \end{subfigure}%
}
    \begin{tabular}{@{}c@{}}
        \usebox{\subcapboxA} \\[2ex]
        \usebox{\subcapboxC}
    \end{tabular}\hfill
    \begin{tabular}{@{}c@{}}
        \usebox{\subcapboxB}
    \end{tabular}

    \caption{(a) Flowchart of the single-shot false-lock detection (FLD) algorithm. (b) Simulated eye diagrams for data transmission at 1~Gb/s through 3~m of radio-pure cable. Top: with default parameter values; bottom: with optimized parameter values. (c) Block diagram of the MATALB-Cadence co-simulation procedure developed for automated optimization of line driver parameters.}
    \label{fig:lock_detector}
\end{figure}

\textbf{False Lock Detector:}~The DLL includes a false lock detector (FLD) circuit that can determine if the final tap is locked to a delay that is not equal to $1\times UI$. This is crucial since if the DLL does not lock correctly, it will degrade the granularity of relative delay options available to the user. While important, the FLD only needs to be run once the system locks. However, earlier FLD designs run continuously, thus wasting power~\cite{razavi_2018}. For this reason, we used the novel one-shot FDL algorithm shown in Fig.~\ref{fig:lock_detector}(a). When run on our 8-stage VCDL, this algorithm compares the relative delays of taps 1-4 and tap 8. Two checks need to be performed during the test, one to determine if there is locking at integer multiples of the UI greater than 1, and another to determine if there is locking at a non-integer multiple of the UI. For the first check, we compare taps 1-4. If the DLL locked correctly, tap 4 would have the maximum relative delay of this set. Therefore, from tap 1 to tap 4, the delay between the tap output and the input clock should increase. If the DLL locks at an integer multiple of the UI greater than 1, the delay will not increase continuously from tap 1 to tap 4. Therefore, we compare the delays of $tap[i]$ and $tap[i+1]$ for $i=\{1,2,3\}$. If any of these comparisons result in $tap[i]$ having a greater or equal rising-edge delay to $tap[i+1]$, then the test fails. If all tests pass, then a final check is performed to detect non-integer locking. This is found by comparing taps 4 and 8. The delay between the final tap and the input clock (modulo $1\times UI$) is ideally zero after locking. However, if the DLL locks on a non-integer value, the delay will be greatest at the final tap. Therefore, we compare taps 4 and 8: if the delay is greater at tap 8, there is a failure, while if the opposite is true the test passes. After detecting a pass or fail, the block shuts down unless initiated to run again by the system, thus conserving power. At the hardware level, the FLD was implemented using a set of modified phase detector (PD) and CP circuits and static logic.

\textbf{Tap Weight Control Circuitry:}~Configurable tap weights require control over the drive strength of each tap output. Voltage-mode source-series terminated (SST) driver arrays~\cite{menolfi200716gb} are used for this purpose. SST drivers are chosen instead of current-mode logic (CML) drivers as they consume 2-$4\times$ less power for the same output swing. Eight parallel SST slices are provided for each of the three taps, and the user can enable any number of slices per tap. Lastly, three different source resistance options are provided, namely 300, 700, and 1100~$\Omega$. This additional flexibility allows one to either improve output impedance matching or maximize/minimize output swing as needed. The schematic of a single SST slice with configurable source resistance is shown in Fig.~\ref{fig:overview}(b).


\section{Output Waveform Optimization}
An automated process was developed for finding the optimum pre-emphasis waveform for a given cable (in this case, the same 3~m long radio-pure cable discussed in Section~1). As shown in Fig.~\ref{fig:lock_detector}(b), the eye diagram can be significantly improved (relative to a conventional FFE with all tap delays equal to $1\times UI$) by also optimizing the tap durations. While this optimum was found via an exhaustive search, a more efficient procedure is essential due to the large amount of available parameter permutations. To this end, we developed a Cadence test bench that contains a Verilog-A model of the line driver and a transistor-level model of the SST arrays driving the chosen cable. This test bench outputs results into MATLAB via a Simulink model and a SpectreRF coupler block that enables bidirectional communication between MATLAB and Cadence using TCP/UDP packets. These results are then fed to a MATLAB optimization function that minimizes a cost function given by $FOM = {\sqrt{I_{av}}}/{(e_{H}\times e_{W})},$
where $I_{av}$ is the average power consumption of the driver and $e_{H}$ and $e_{W}$ are the height and width of the eye opening with random input data, respectively. These parameters are combined into a single Figure of Merit (FOM) that is designed to favor a reasonable compromise between power consumption (proportional to $I_{av}$) and BER (strongly decreasing with $e_H$ and $e_W$).
Fig.~\ref{fig:lock_detector}(c) illustrates the entire co-simulation and optimization flow.

\section{Test Structure and Results}
\begin{figure}
    \begin{subfigure}{0.26\textwidth}
        \centering
        \includegraphics[width=1\textwidth]{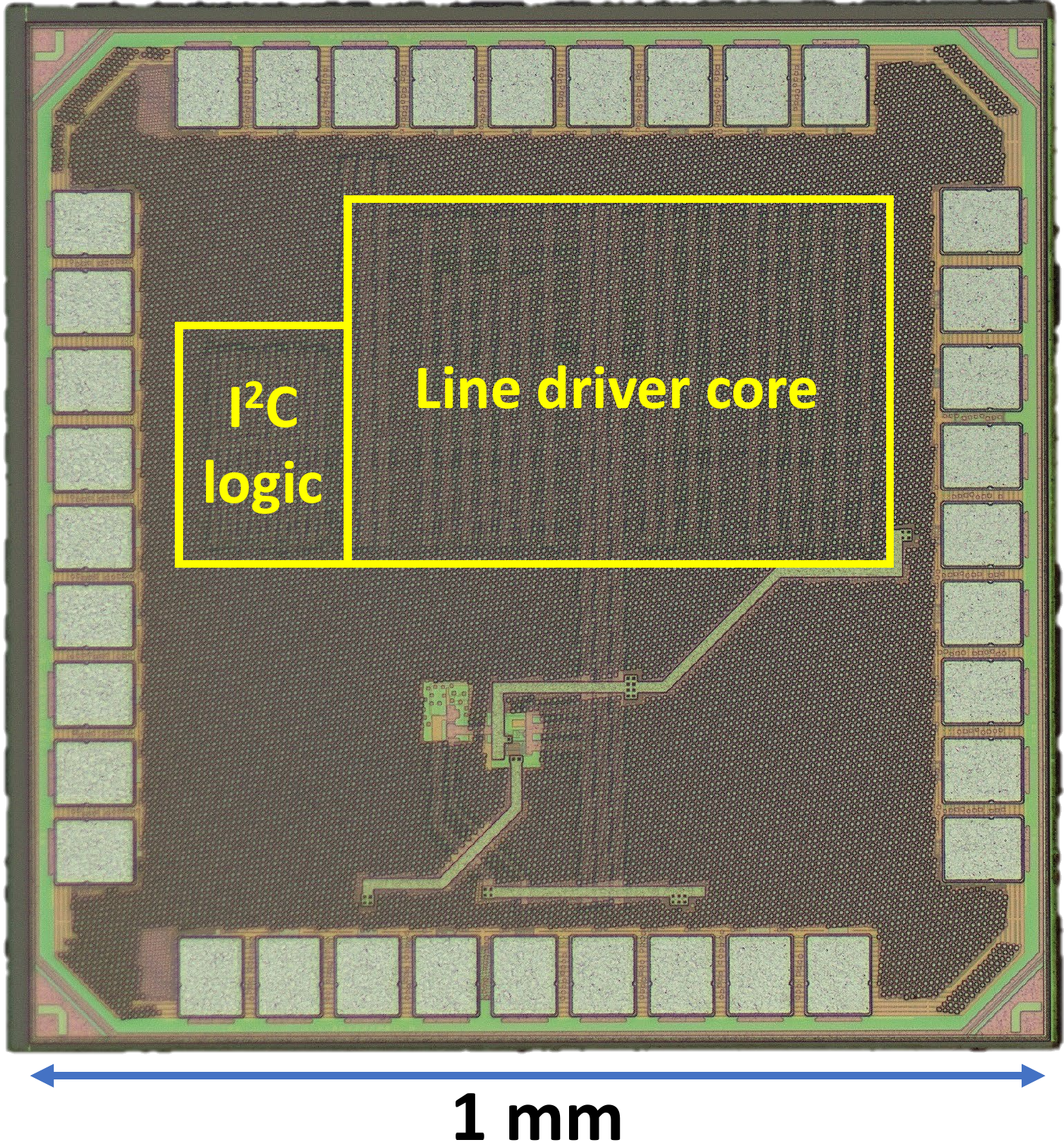}
        \caption{}
    \end{subfigure}~
    \begin{subfigure}{0.71\textwidth}
        \centering
        \includegraphics[width=1\textwidth]{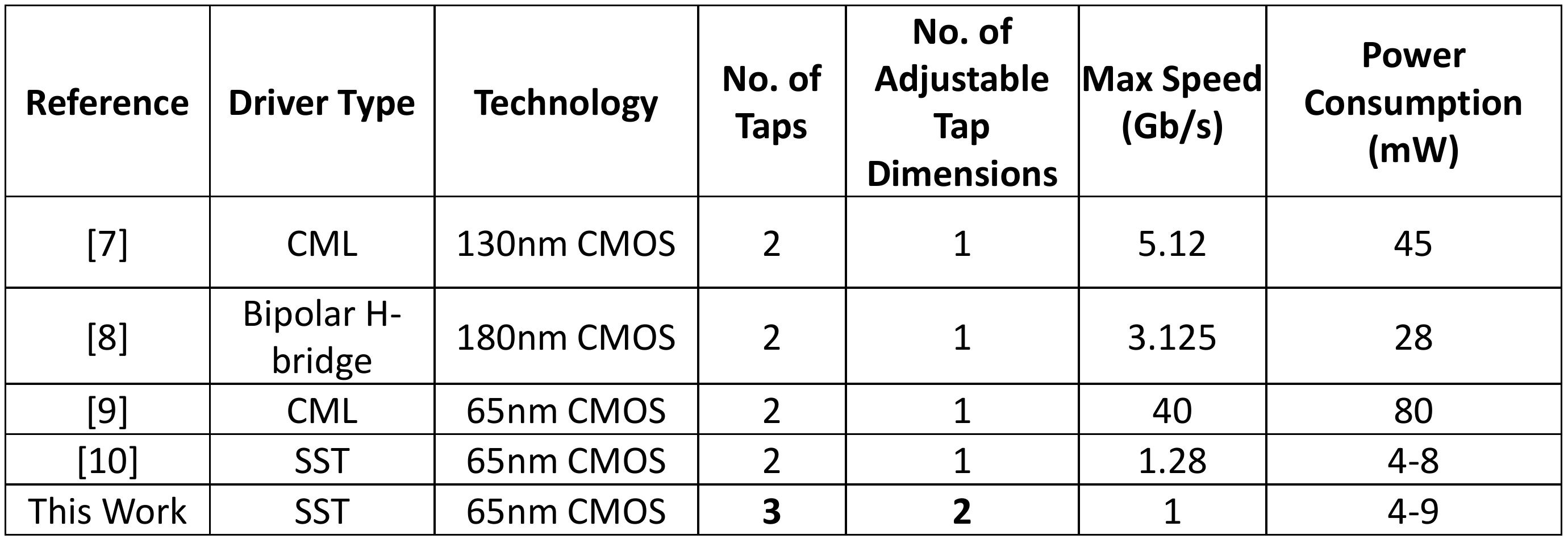}
        \caption{}
    \end{subfigure}
    \caption{(a) Labeled die photograph of the test chip, which has a total area of 1~mm$^2$. (b) Comparison of line driver performance with comparable designs in the literature.}
    \label{fig:expt_results}
\end{figure}

Fig.~\ref{fig:expt_results}(a) shows a die photograph of the fabricated test chip, which integrates the line driver with various clocking and data generation options and an I$^{2}$C control interface. The line driver core occupies $355\times 363$~$\mu$m. Total power consumption varies from 4-9~mW while the maximum data rate is limited to 1~Gb/s by the DLL design. The design is compared with other work~\cite{gromov2015development,iniewski20053,huang201580,Moreira:2809058} in Fig.~\ref{fig:expt_results}(b), which shows that our work maximizes user configurability while minimizing power. 

\section{Conclusion}
We have described a highly-configurable line driver circuit in 65~nm CMOS that is particularly suitable for driving high-loss channels. The circuit uses a 3-tap FFE with fine-grained control over both tap weights and time delays, thus allowing the user to adapt the output waveform for a variety of cable types and data rates. Future work will focus on i) higher data-rate implementations, ii) replacing the existing DLL with an all-digital version, and iii) implementing self-monitoring algorithms to enable the driver to autonomously optimize its performance over time.

\bibliographystyle{JHEP}
\bibliography{references.bib}

\end{document}